\documentclass[
 reprint,
 amsmath,amssymb,superscriptaddress,
 aps,
 prl,
]{revtex4-1}

\usepackage{graphicx}
\usepackage{dcolumn}
\usepackage{bm}
\usepackage{color}
\usepackage{ulem}
\usepackage{float}
\usepackage{multirow}

\restylefloat{table}

\begin{document}

\title{Determination of the Fermi Contour and Spin-polarization of $\nu=3/2$ Composite Fermions via Ballistic Commensurability Measurements}
\date{\today}

\author{D.\ Kamburov}
\author{M. A.\ Mueed}
\author{I.\ Jo}
\author{Yang \ Liu}
\author{M.\ Shayegan}
\author{L. N.\ Pfeiffer}
\author{K. W.\ West}
\author{K. W.\ Baldwin}
\author{J. J. D.\ Lee}
\affiliation{ Department of Electrical Engineering, Princeton University, Princeton, New Jersey 08544, USA}

\author{R.\ Winkler}
\affiliation{Department of Physics, Northern Illinois University, DeKalb, Illinois 60115, USA}
\affiliation{Materials Science Division, Argonne National Laboratory, Argonne, Illinois 60439, USA}

\begin{abstract}

We report ballistic transport commensurability minima in the magnetoresistance of $\nu =3/2$ composite fermions (CFs). The CFs are formed in high-quality two-dimensional electron systems confined to wide GaAs quantum wells and subjected to an in-plane, unidirectional periodic potential modulation. We observe a slight asymmetry of the CF commensurability positions with respect to $\nu=3/2$, which we explain quantitatively by comparing three CF density models and concluding that the $\nu=3/2$ CFs are likely formed by the minority carriers in the upper energy spin state of the lowest Landau level. Our data also allow us to probe the shape and size of the CF Fermi contour. At a fixed electron density of $\simeq 1.8 \times 10^{11}$ cm$^{-2}$, as the quantum well width increases from 30 to 60 nm, the CFs show increasing spin-polarization. We attribute this to the enhancement of the Zeeman energy relative to the Coulomb energy in wider wells where the latter is softened because of the larger electron layer thickness. The application of an additional parallel magnetic field ($B_{||}$) leads to a significant distortion of the CF Fermi contour as $B_{||}$ couples to the CFs' out-of-plane orbital motion. The distortion is much more severe compared to the $\nu=1/2$ CF case at comparable $B_{||}$. Moreover, the applied $B_{||}$ further spin-polarizes the $\nu=3/2$ CFs as deduced from the positions of the commensurability minima.  

\end{abstract}


\maketitle

\section{I. Introduction}

The composite fermion (CF) formalism offers an elegant explanation for the physics of interacting two-dimensional electron systems (2DESs) at very low temperatures in the presence of a strong perpendicular magnetic field $(B_{\perp})$. In the CF framework, the electron-electron interaction manifests itself through the formation of CFs, quasi-particles resulting from the attachment of an even number of flux quanta to each electron at high $B_{\perp}$ \cite{Jain.2007,Jain.PRL.1989,Halperin.PRB.1993}. At even-denominator Landau level (LL) filling factors, e.g. at $\nu=1/2$ and 3/2, the flux attachment cancels the external field completely, leaving the CFs as if they are at zero effective magnetic field $B^*$. Analogous to their low-field counterparts, near these fillings the CFs occupy a Fermi sea with a well-defined Fermi contour \cite{Jain.2007,Jain.PRL.1989,Halperin.PRB.1993,Willett.PRL.1993,Kang.PRL.1993,Goldman.PRL.1994,Smet.PRL.1996,Endo.PRB.2001,Kamburov.CFs.2012,hCFanisotropy.Kamburov.2012,eCFanisotropy.Kamburov.2012,Smet.PRL.1998,Willett.PRL.1997,Oppen.PRL.1998,Smet.PRL.1999}. 

At $\nu=3/2$, the lower energy spin state of the lowest $(N=0)$ LL, the $|0\uparrow\rangle$ state, is fully occupied while its higher energy spin state, $|0\downarrow\rangle$, is half filled. In a simple picture, one might expect the $\nu=3/2$ CFs to be fully spin-polarized. However, the CF spin-polarization depends on the interplay between the Coulomb energy $(E_C=e^2/4 \pi \epsilon l_B)$, and the Zeeman energy $(E_Z=\mu_B gB)$, where $l_B=(\hbar/eB)^{1/2}$ is the magnetic length, $\epsilon$ is the dielectric constant, $g$ is the Lande $g$-factor, $e$ is the electron charge, and $\mu_B$ is the Bohr magneton  \cite{Clark.PRL.1989,Eisenstein.PRL.1989,Engel.PRB.1992,Jain.Park.PRL.1998,Kukushkin.PRL.1999,Davenport.PRB.2012,Vanovsky.PRB.2013,gfactor,Yang.PRB.2014,Du.PRL.1995}. When $E_C$ fully dominates over $E_Z$, the CFs can be partially spin-polarized. This is indeed evidenced by transitions observed between fractional quantum Hall states with different spin-polarizations as $E_Z/E_C$ is varied \cite{Clark.PRL.1989,Eisenstein.PRL.1989,Engel.PRB.1992,Jain.Park.PRL.1998,Kukushkin.PRL.1999,Davenport.PRB.2012,Vanovsky.PRB.2013,Yang.PRB.2014,Du.PRL.1995}. The CF spin-polarization near $\nu=3/2$ can also be deduced directly from measurements of the size of the CF Fermi wave vector $k_F^*$. However, such experiments have been elusive \cite{Endo.PRB.2001}. Here we report direct measurements of the $\nu=3/2$ CF Fermi contour size and shape using commensurability oscillations. We also deduce the CF spin-polarization and present three models to treat the non-trivial question of the number of CFs in the vicinity of $\nu=3/2$.

\section{II. Overview of Results}

We perform $\nu=3/2$ CF commensurability measurements in very high-quality 2DESs confined to GaAs quantum wells (QWs) of widths $W$= $30$ to $60$ nm. The 2DESs have a surface-strain-induced, unidirectional, density modulation. In particular, we monitor, near $\nu=3/2$, the positions of $B^*$ at which we observe magnetoresistance features. Associating the features with the commensurability between the majority-spin CF cyclotron orbit and the period of the density modulation, our data are consistent with the CFs occupying a circular Fermi sea \cite{majority_spin_CF}.

The positions of the commensurability features are slightly asymmetric in magnetic field with respect to $\nu=3/2$. The asymmetry, best accounted for by a \textit{minority-density} model in the $|0\downarrow\rangle$ LL, in which the CFs are formed by \textit{electrons} for $\nu<3/2$ and by \textit{holes} for $\nu>3/2$, is similar to an asymmetry observed in the commensurability oscillations of $\nu=1/2$ CFs \cite{Kamburov.2014}. It indicates a subtle breaking of the particle-hole equivalence in the ballistic regime of CFs and points to the nontrivial and fundamental question which particles, electrons or holes, pair with flux quanta to form CFs in the vicinity of even-denominator filling factors.

The Fermi contour area we deduce from the positions of the commensurability features implies that the $\nu=3/2$ CFs' degree of spin-polarization depends on the electron density $n$ and on the width $W$ of the QW in which the CFs are confined. In narrower QWs and at low densities, the CFs are partially spin-polarized, but at higher densities, or when $W=60$ nm, they reach full spin-polarization. We explain the $W$- and $n$-dependence of the degree of CF spin-polarization based on the fact that higher density leads to an increase in $E_Z/E_C$ and a wider QW reduces $E_C$. Our measurements also reveal that the commensurability features are more pronounced when the CFs are more spin-polarized. 

We further apply a magnetic field ($B_{||}$) parallel to the 2D plane. Because of the finite thickness of the electron layer, $B_{||}$ couples to the carriers' out-of-plane orbital motion and severely distorts the $\nu=3/2$ CF Fermi contour. Remarkably, this $B_{||}$-induced distortion is much more pronounced compared to the CF Fermi contour anisotropy observed near $\nu=1/2$ at a comparable $B_{||}$ \cite{eCFanisotropy.Kamburov.2012,hCFanisotropy.Kamburov.2012}. In addition, thanks to the increasing $E_Z$, which depends on the total magnetic field, the degree of spin-polarization of the CFs increases with $B_{||}$.

\section{III. Experimental details}

The 2DESs in our samples are confined to symmetric GaAs QWs grown via molecular beam epitaxy on (001) GaAs substrates. We studied five samples with QW widths $W$ of 30, 40, 50, and 60 nm. Four of the QWs are 190 nm under the surface and are flanked on each side by 150-nm-thick, undoped Al$_{0.24}$Ga$_{0.76}$As barrier (spacer) layers and Si $\delta$-doped layers. After a brief illumination with a red light emitting diode at low temperatures, the 2D electron densities in these samples are $n \simeq 1.7-1.9 \times 10^{11}$ cm$^{-2}$. The fifth sample, another 40-nm-wide QW located 135 nm under the surface with a 95-nm-thick spacer, has $n \simeq 2.7 \times 10^{11}$ cm$^{-2}$. The mobilities of all samples are $\simeq 10^7$ cm$^2$/Vs. We did experiments in two pumped $^3$He cryostats with base temperatures $T \simeq 0.3$ K and with either an 18 T superconducting or a 31 T resistive magnet. The magnitude of $B_{||}$ was varied by tilting the samples \textit{in situ} around $[\overline{1}10]$ so that $B_{||}$ was always along the $[110]$ direction, with $\theta$ denoting the angle between the field direction and the normal to the 2D plane (see Fig. 1 inset). In our measurements each sample has two perpendicular Hall bars, along the $[110]$ and $[\overline{1}10]$ directions, covered with periodic gratings of negative electron-beam resist with a period $a = 200$ nm (see Fig. 1 inset). The stripes impart strain to the sample surface which, through the piezoelectric effect in GaAs, induces a periodic 2DES density modulation \cite{Endo.PRB.2001,eCFanisotropy.Kamburov.2012,hCFanisotropy.Kamburov.2012,Kamburov.PRB.2012b,Kamburov.CFs.2012,Skuras.APL.1997,Endo.PRB.2000,Kamburov.2014}. In the presence of an applied $B_{\perp}$, whenever the diameter of the CFs' cyclotron orbit becomes commensurate with the period of the density modulation, the sample's resistance exhibits a minimum.

\section{IV. Commensurability features of $\nu=3/2$ composite fermions}

In the CF formalism, away from $\nu=3/2$, the quasi-particles feel an effective magnetic field $B^*$ and undergo cyclotron motion with radius $R^*_C=2 \hbar k^*_F/eB^*$, where $k_F^*=\sqrt{4 f \pi n^*}$ is the CF Fermi wave vector, $n^*$ is the CF density, and $f$ is the degree of spin-polarization ($f=0.5$ means the CFs are spin-unpolarized and $f=1$ signifies full spin-polarization). The magnitude of $B^*$ should depend on the number of electrons in the $|0\downarrow\rangle$ state, which, in a simple picture, can be approximated by $n/3$. Consequently, the CF density is $n^*=n/3$, and the effective magnetic field experienced by the CFs is $B^*=3(B-B_{3/2})$, where $B_{3/2}$ is the magnetic field at $\nu=3/2$. The periodic surface strain in our samples induces a local variation of the CF density $n^*$ which in turn creates a strong local variation of $B^*$. In Fig. 1 we show the magnetoresistance traces at tilt angle $\theta = 0^{\circ}$ ($B_{||}=0$) along the two Hall bar arms of the 40-nm-wide QW sample with $n=2.71 \times 10^{11}$ cm$^{-2}$. They exhibit prominent commensurability features near $\nu = 3/2$, including a characteristic, V-shaped, resistance dip, followed by resistance minima on the sides of $\nu=3/2$ and flanked by regions of rapidly rising resistance. This behavior is qualitatively similar to the commensurability features near $\nu=1/2$ \cite{Endo.PRB.2001,Kamburov.CFs.2012,eCFanisotropy.Kamburov.2012,hCFanisotropy.Kamburov.2012,Willett.PRL.1997,Smet.PRL.1998,Oppen.PRL.1998,Smet.PRL.1999,Kamburov.2014}. Commensurability features are absent in the trace from the unpatterned region, shown with dashed line in Fig. 1.

In principle, periodic variations in both $n^*$ and $B^*$ can trigger commensurability effects of their own. Magnetoresistance minima are thus anticipated when the cyclotron radius $R^*_C$ of CFs' quasi classical orbits satisfies the condition \cite{Kamburov.CFs.2012,eCFanisotropy.Kamburov.2012,hCFanisotropy.Kamburov.2012,Willett.PRL.1997,Smet.PRL.1998,Oppen.PRL.1998,Smet.PRL.1999,Kamburov.2014,Endo.PRB.2001}:
\begin{equation} 2R_C^*/a=i \pm 1/4,
\label{eq:2}
\end{equation}
where $i$ is a positive integer. The negative sign denotes the \textit{electrostatic} commensurability condition, corresponding to a variation in $n^*$, while the positive sign implies that the commensurability is \textit{magnetic}, i.e. it is related to a periodic modulation of $B^*$. Previous experiments near $\nu=1/2$ reveal excellent agreement between the CF commensurability minima and the \textit{magnetic} condition, indicating the CFs there experience a modulated effective field $B^*$ \cite{Kamburov.CFs.2012,eCFanisotropy.Kamburov.2012,Endo.PRB.2001,hCFanisotropy.Kamburov.2012,Willett.PRL.1997,Smet.PRL.1998,Oppen.PRL.1998,Smet.PRL.1999,Kamburov.2014}.

\begin{figure}[!t]
\includegraphics[trim=1.2cm 0.2cm 0cm 0cm, clip=true, width=0.48\textwidth]{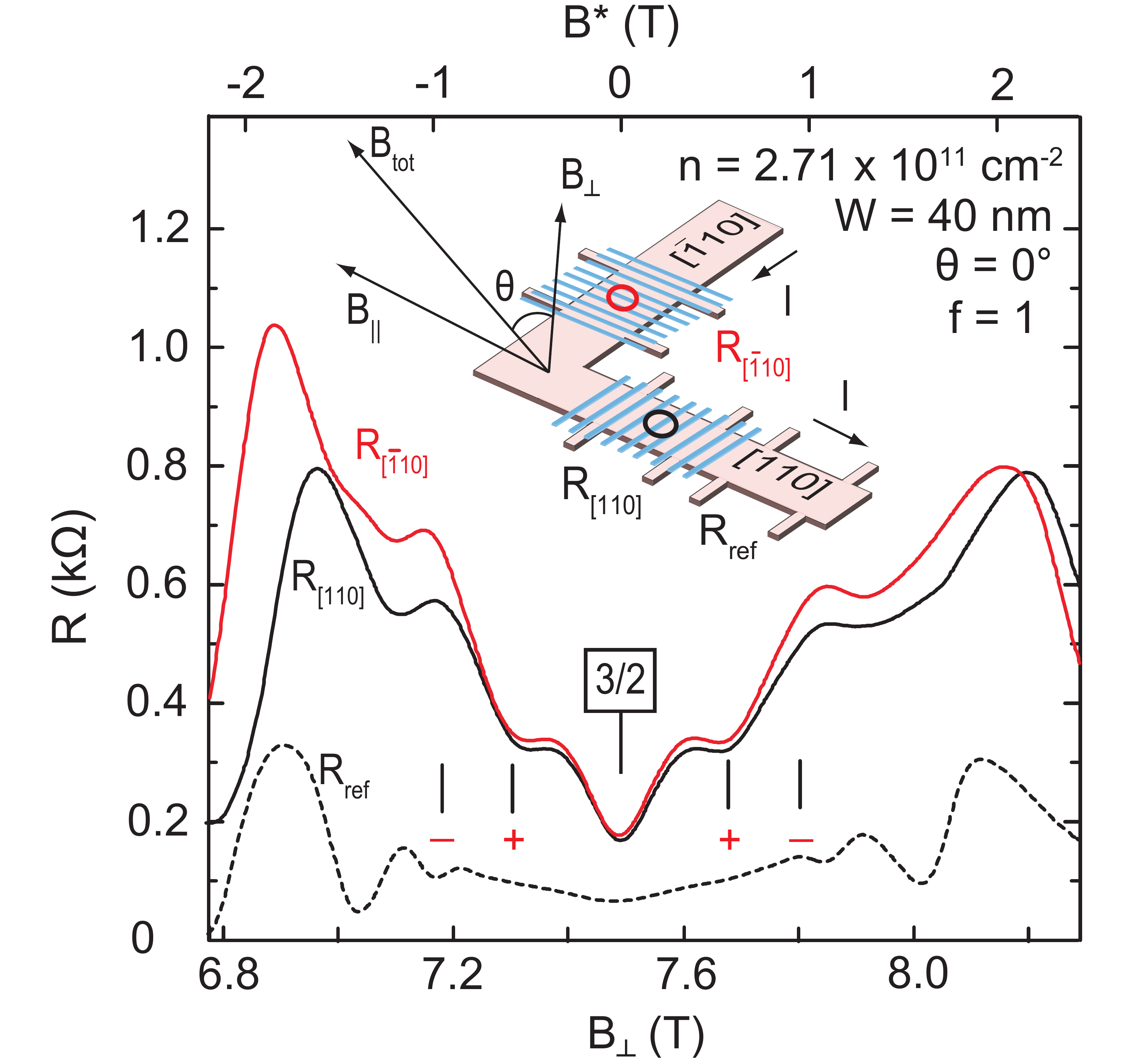}
\caption{\label{fig:Fig1} (color online) Inset: Schematic of the L-shaped Hall bar showing the regions covered with periodic gratings of negative electron-beam resist. The parallel magnetic field $B_{||}$ is applied along the $[110]$ direction by tilting the sample to angle $\theta$ with respect to the total magnetic field $B_{\textbf{tot}}$. Main: Magnetoresistance traces (solid lines) measured at $\theta = 0^{\circ}$ in the patterned sections of the L-shaped Hall bar along $[\overline{1}10]$ and $[110]$, showing strong commensurability features in the vicinity of $\nu=3/2$. Commensurability features are absent in the reference region data (dashed trace). Two sets of vertical lines mark the expected resistance minima for magnetic $(+)$ and electrostatic $(-)$ commensurability based on Eq. (1) with $n^*=n/3$, $f=1$, and $B^*=3(B-B_{3/2})$, where $B_{3/2}$ is the field at $\nu=3/2$. The observed minima agree with the \textit{magnetic} commensurability condition.}
\end{figure}

\begin{figure}[!t]
\includegraphics[trim=0.8cm 0.4cm 1.6cm 0.2cm, clip=true, width=0.45\textwidth]{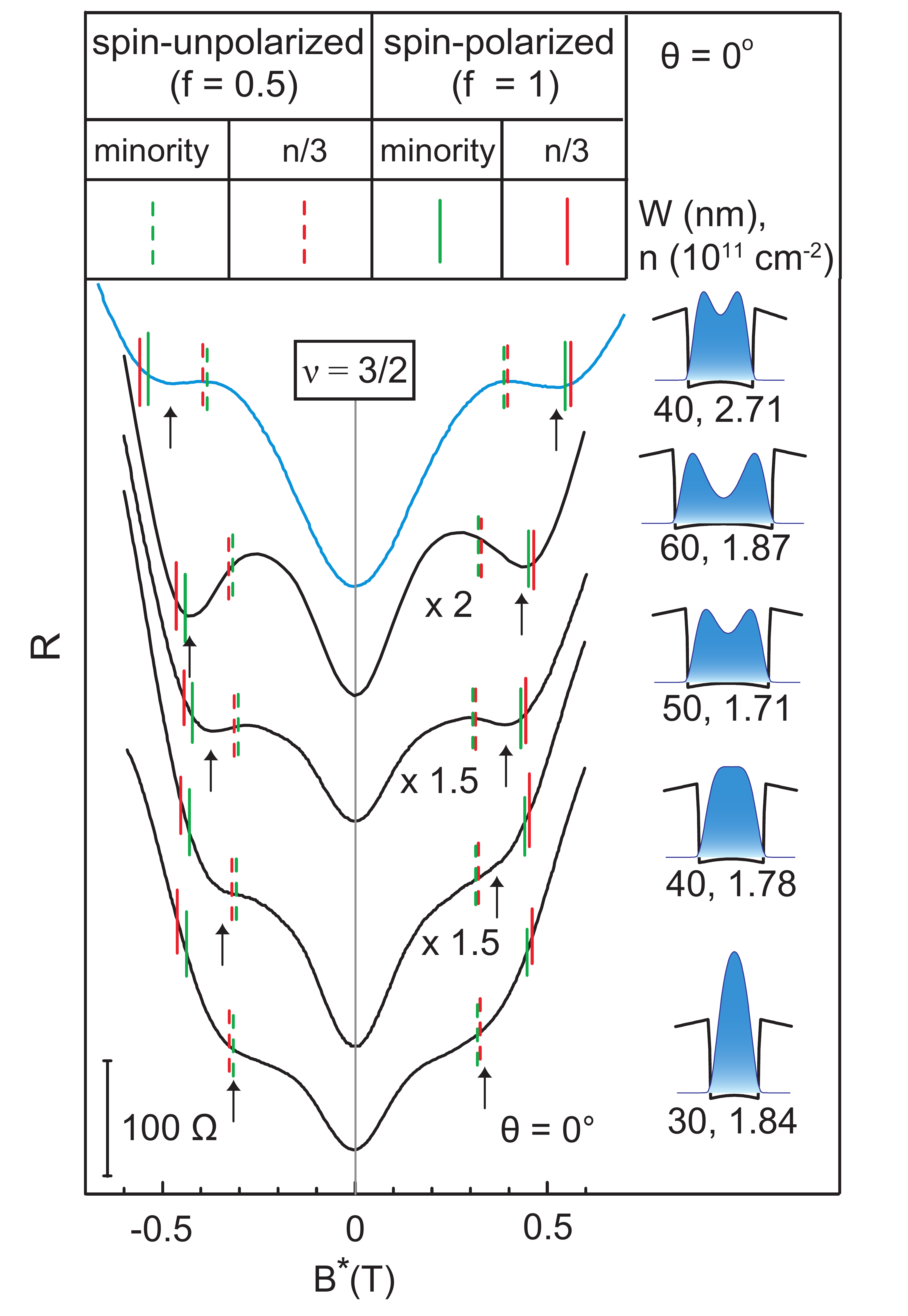}
\caption{\label{fig:Fig2} (color online) Magnetoresistance traces from all five samples taken at tilt angle $\theta=0^{\circ}$. The top trace, shown in blue, has a much higher density than the other traces. All traces exhibit commensurability features whose positions are marked by black vertical arrows. Also shown for each sample are different sets of vertical marks, indicating the expected positions of the $i=1$ commensurability resistance minimum, based on different assumptions. The red marks assume that the CF density is $n^*=n/3$ while the green lines are based on the assumption that $n^*$ equals the density of minority carriers in the $|0\downarrow\rangle$ LL (see Table I). Solid lines are for full spin-polarization $(f=1)$ and dashed ones are for the spin-unpolarized case $(f=0.5)$. A self-consistent calculation of the charge distribution and the potential profile for the samples is shown to the right of each trace. }
\end{figure}

Quantitatively, the positions of the resistance minima we observe near $\nu=3/2$ agree well with the commensurability condition for CFs with a circular Fermi contour in a weak periodic \textit{magnetic} field, similar to the $\nu=1/2$ CFs. In Fig. 1, two sets of vertical lines, symmetric around $\nu=3/2$, mark the expected positions of the CF commensurability minima based on Eq. (1) applied to a simple picture in which $n^*=n/3$ and $B^*=3(B-B_{3/2})$ with $f=1$ and $i=1$. The lines closer to $\nu=3/2$ correspond to the magnetic condition $(+)$, and those further away mark the expected electrostatic commensurability minima $(-)$. Clearly the lines marked with $+$ match the observed resistance minima, providing evidence that the data are consistent with the magnetic commensurability condition. We conclude that in our samples a surface-strain-induced density modulation directly affects $B^*$ near $\nu=3/2$ and makes it spatially periodic.  

\begin{table*}[!t]

\begin{center}
\renewcommand{\arraystretch}{1.6}

\scalebox{1.0}{

\begin{tabular}{| >{\centering\arraybackslash}m{0.5in}| >{\centering\arraybackslash}m{1in}| >{\centering\arraybackslash}m{1.2in}| >{\centering\arraybackslash}m{1.5in}| } \\ [-6ex]

\hline\hline

Model   &  $n^*$ & validity & position\\\hline
\multicolumn{1}{|c|}{\multirow{2}{*}{fixed-density}} & \multicolumn{1}{c|}{\multirow{2}{*}{n/3}} & $B>B_{3/2}$ & $B^*=B'^*_>=\frac{2\hbar\sqrt{4 \pi fn/3}}{ea(1+i/4)}$\\[1ex]\cline{3-4}
\multicolumn{1}{|c|}{} & \multicolumn{1}{c|}{} & $B<B_{3/2}$ & $B^*=B'^*_<=- \frac{2\hbar\sqrt{4 \pi fn/3}}{ea(1+i/4)}$\\[1ex]\hline

\multicolumn{1}{|c|}{\multirow{2}{*}{minority-density}}   &  $\frac{\nu-1}{\nu}n$ & $B>B_{3/2}$ & $B^{*} \simeq B'^*_>-\frac{(B'^{*})^2}{3B_{3/2}}$\\\cline{2-4}
\multicolumn{1}{|c|}{}   &  $\frac{2-\nu}{\nu}n$ & $B<B_{3/2}$ & $B^{*} \simeq B'^*_<+\frac{2(B'^{*})^2}{3B_{3/2}}$\\\hline

\multicolumn{1}{|c|}{\multirow{2}{*}{hole-density}} & \multicolumn{1}{c|}{\multirow{2}{*}{$\frac{2-\nu}{\nu}n$}} & $B>B_{3/2}$ & $B^{*} \simeq B'^*_>+\frac{2(B'^{*})^2}{3B_{3/2}}$\\[1ex]\cline{3-4}
\multicolumn{1}{|c|}{} & \multicolumn{1}{c|}{} & $B<B_{3/2}$ & $B^{*} \simeq B'^*_<+\frac{2(B'^{*})^2}{3B_{3/2}}$\\[1ex]\hline
\end{tabular}

}
\end{center}
\caption{Summary of the density models for CF density $n^*$ near $\nu=3/2$. The fixed-density model assumes the CF density equals the electron density. In the minority-density model, $n^*$ is taken to be the density of holes for $B<B_{3/2}$ and of electrons when $B>B_{3/2}$. The hole-density model assumes $n^*$ is equal to the density of holes in the $|0\downarrow\rangle$ LL. \textcolor{white}{ffffffffffffffffffffffffffffffffffffffffffffffffffffffffffffffffffffffffffffffffffffffffffffffffffffffffffffffffffffffffffffffffffffffffffffffffffffffffffffffffffffffffffffffffffffffffffffffffff}}
\label{table:CFdensity}
\end{table*}

The results of our measurements at $\theta=0^{\circ}$ are summarized in Fig. 2, where we show one trace from a patterned section for every sample. The corresponding QW widths and densities are also indicated along with a calculation of the charge distribution in the absence of magnetic field. The density of the top trace, shown in Fig. 2 in blue, is significantly higher compared to the other traces. In the vicinity of $\nu=3/2$, each trace exhibits commensurability features. We mark their positions with vertical arrows. The features in the $W=30$ nm and the lower-density $W=40$ nm samples (bottom two traces in Fig. 2) are rather weak. As the QW width and/or the electron density increase, the features develop into pronounced resistivity minima and move outward from $\nu=3/2$. The positions of the minima are slightly asymmetric with respect to $\nu=3/2$ with those on the $B^*<0$ side being closer to $\nu=3/2$. 

\begin{figure}[!b]
\includegraphics[trim=0cm 0cm 0cm 0.8cm, clip=true, width=0.48\textwidth]{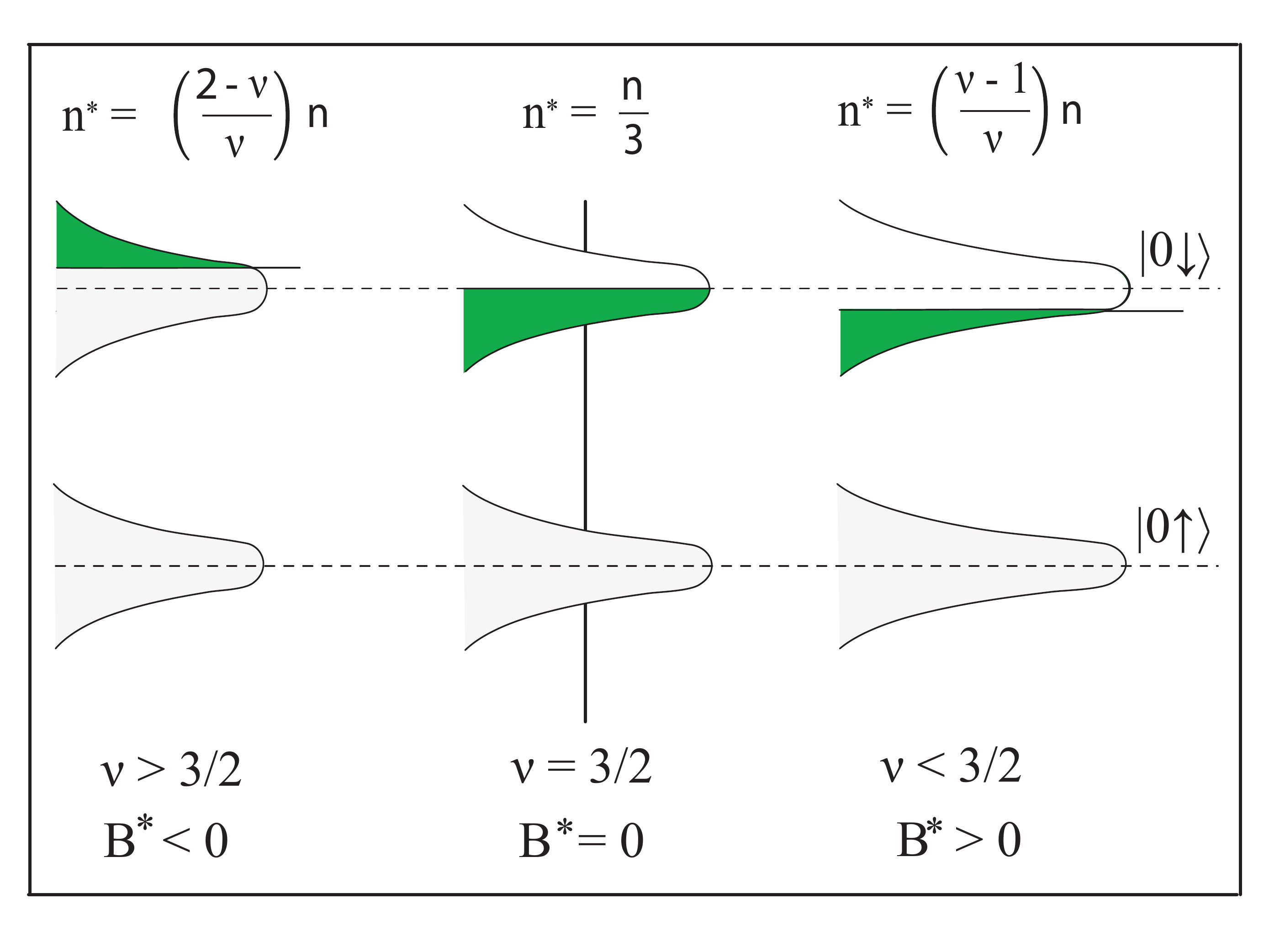}
\caption{\label{fig:Fig3} (color online) Schematic of the LL population in the vicinity of $\nu=3/2$ in the context of the minority-density model, which assumes that the CF density $n^*$ equals the density of the minority carriers (shaded in green) in the $|0\downarrow\rangle$ LL, i.e. $n^*$ is the density of \textit{electrons} for $\nu<3/2$ and of \textit{holes} for $\nu>3/2$.}
\end{figure}

In order to quantitatively understand the positions of the commensurability features and their dependence on the QW width and density, we compare them to the expected values based on the commensurability between the quasi-classical orbits of the CFs and the periodic potential modulation (Eq. (1) with a positive sign). For such a comparison, we need to treat the CF density $n^*$ explicitly in order to predict the $B^*$ values at the commensurability condition. Determining the CF density is a non-trivial problem because, away from $\nu=3/2$, the electrons and holes in the $|0\downarrow\rangle$ LL have unequal densities. It is similar to an earlier study which revealed an unexpected asymmetry in the positions of the CF commensurability features around $\nu=1/2$, both as a function of field and filling factor \cite{Kamburov.2014}. The asymmetry there was explained by assuming that the CF density is equal to the density of the minority carriers in the spin-resolved, lowest LL, namely, the CF density was taken to be the density of electrons when $\nu < 1/2$ but of holes when $\nu > 1/2$. The results near $\nu=1/2$ suggest that CFs are formed by pairing of flux quanta with the minority carriers in the lowest LL and further indicate a breaking of the particle-hole symmetry \cite{Kamburov.2014}.

\begin{figure*}[t!]
\includegraphics[trim=0cm 0.8cm 0.0cm 0cm, clip=true, width=0.98\textwidth]{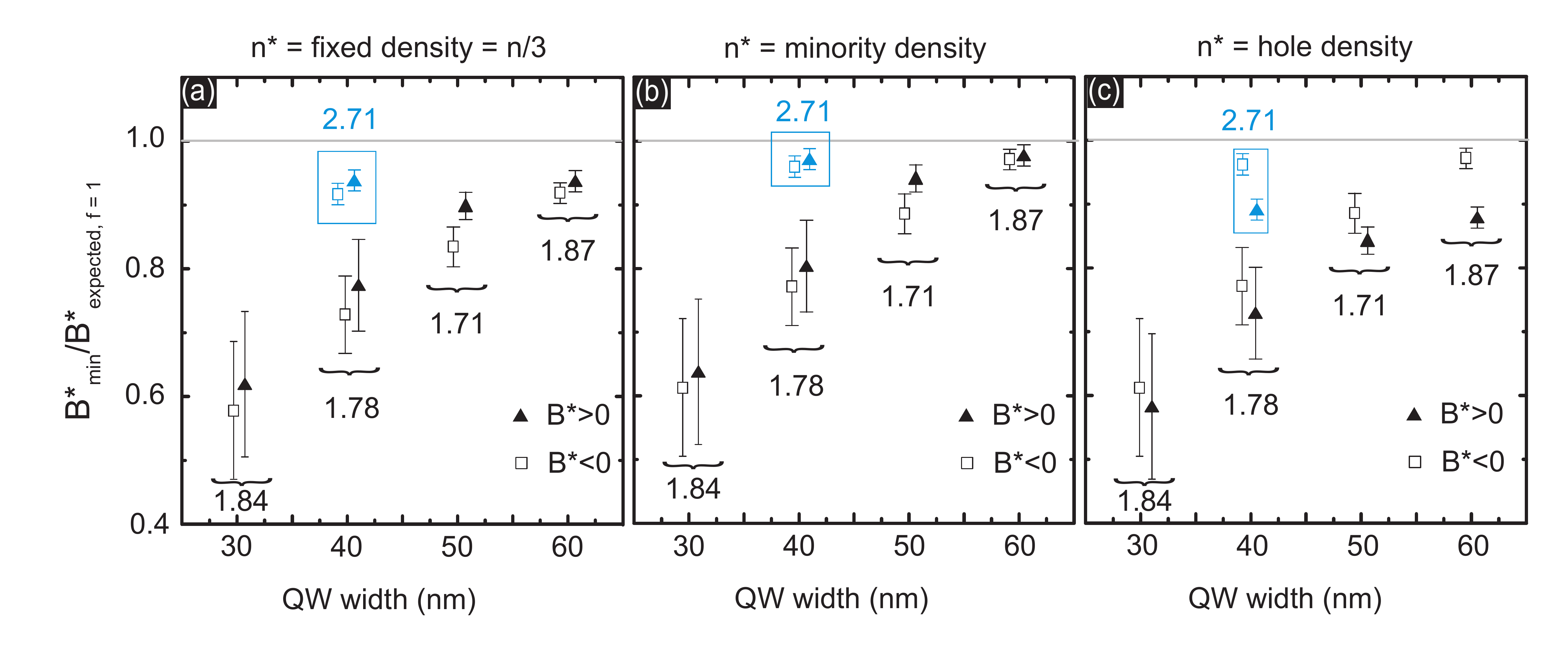}
\caption{\label{fig:Fig4} (color online) Observed $B^*$ values of the commensurability resistance minima or features, normalized to the expected positions based on Eq. (1), and the CF density $n^*$. In (a) $n^*=n/3$; (b) treats CFs as formed by the minority carriers in the $|0\downarrow\rangle$ LL; (c) shows the normalized values if $n^*$ equals the density of holes in the same level. Representative error bars are shown for each data set. Data for $B^*>0$ are shown using closed triangles and, for $B^*<0$, using open squares. The sample density is given in units of $10^{11}$ cm$^{-2}$ for each QW and, for clarity, data for a given QW width $W$ are shown slightly displaced horizontally to avoid overlap of the error bars. Each data point is an average of two points from the traces taken along [110] and $[\overline{1}10]$. In the $W=30$ nm and the lower density $W=40$ nm samples, where the commensurability features are not clear minima, the error bars are much larger. The positions of $B^*$ in those samples were extracted from the derivative of each trace. }
\end{figure*}

In the present study we consider in detail three density models whose expected resistance minima positions on the two sides of $\nu=3/2$ are summarized in Table I. The first CF density model, which we refer to as the "fixed-density" model, assumes that CFs have a fixed density, $n^*=n/3$, equal to the electron density in the $|0\downarrow\rangle$ LL at $\nu=3/2$. In this case the solution to the magnetic commensurability condition from Eq. (1), when $i=1$, is $B'^* = \pm 2 \hbar \sqrt{4 f \pi n/3}/ea(5/4)$. The two roots, $B'^*_>$ and $B'^*_<$, are symmetric around $\nu=3/2$ and are shown with vertical red lines in Fig. 2. In the second "minority-density" model, we take into account the variation of the electron density in the $|0\downarrow\rangle$ LL away from $\nu=3/2$. We assume that CFs are formed by the \textit{minority} carriers in the $|0\downarrow\rangle$ LL, shaded green in Fig. 3, i.e. by holes for $B<B_{3/2}$ and by electrons for $B>B_{3/2}$. Hence, the CF density is given by $n^*=[(2-\nu)/\nu]n$ for $B<B_{3/2}$ ($\nu>3/2$) and $n^*=[(\nu-1)/\nu]n$ for $B>B_{3/2}$ ($\nu < 3/2$). The solutions to the commensurability condition (Eq. (1)) are \textit{asymmetric} about $\nu=3/2$ (see Table I). They are marked with vertical green lines in Fig. 2. Compared to the fixed-density model, the expected minima in this case are closer to $\nu=3/2$. We also consider a third "hole-density" model (not shown in Fig. 2), where the CFs are assumed to be formed by holes in the $|0\downarrow\rangle$ LL on both sides of $\nu=3/2$. This model was used in Ref. \cite{Endo.PRB.2001} to discuss CF commensurability features near $\nu=3/2$. In this model, the expected positions of $B^*$ are also asymmetric with respect to $\nu=3/2$ (see Table I). On the $B<B_{3/2}$ side, since holes are the minority carriers in the $|0\downarrow\rangle$ LL, the solution is exactly the same as in the minority-density model. On the $B>B_{3/2}$ side, however, the expected solution in the hole-density model is at even higher $B^*$ than the fixed CF density solution. As we discuss later in the context of CF spin-polarization, this model appears inconsistent with our data.

\begin{figure*}[t]
\includegraphics[trim=0.4cm 0.4cm 0.4cm 0.0cm, clip=true, width=0.99\textwidth]{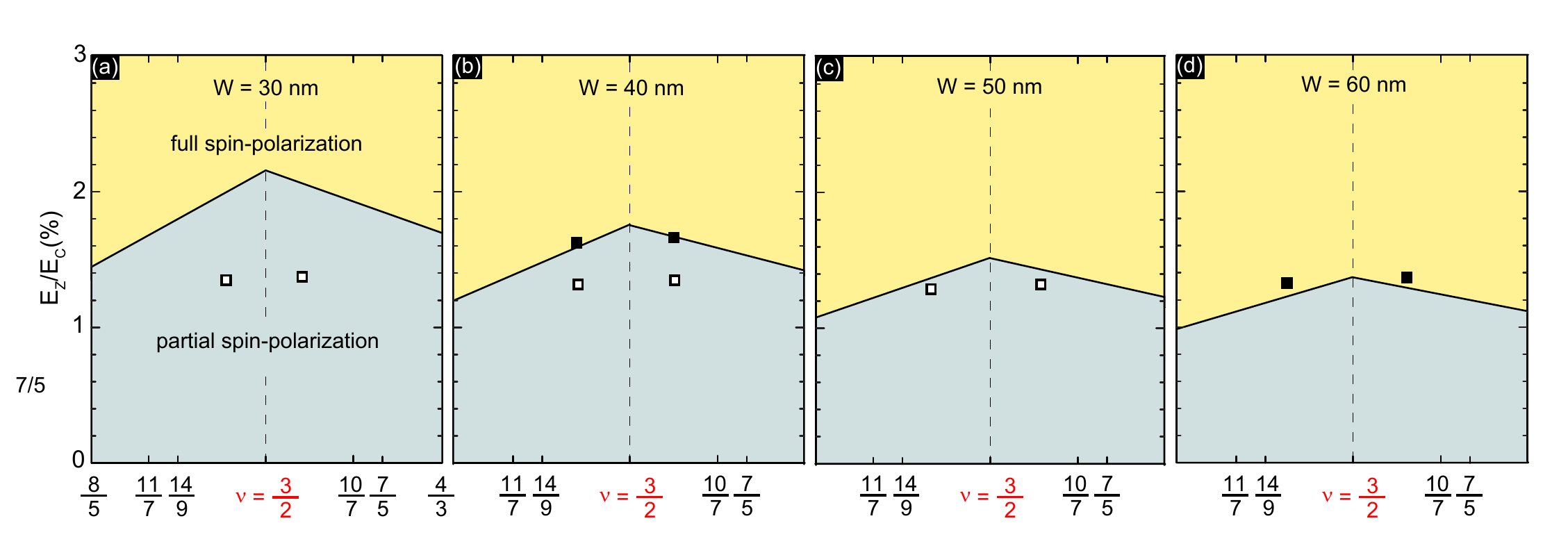}
\caption{\label{fig:Fig5} (color online) In each diagram, symbols represent the experimentally measured ratio of the Zeeman to the Coulomb energies, $E_Z/E_C$, from the field positions of the commensurability features. They are compared to the boundary, shown with solid lines, above which the CFs are observed to become fully spin-polarized based on the data in Ref. \cite{Yang.PRB.2014}. The data for the high-density $W=40$ nm and the 60 nm QWs, represented with closed symbols, are consistent with full spin-polarization. The data from the other samples, marked with open symbols, suggest partial spin-polarization. }
\end{figure*}

The different CF density models predict positions of $B^*$ which are very close to each other (Fig. 2). Since the traces from the narrower QWs exhibit only broad commensurability features rather than clear minima, it is difficult to discern which model is most consistent with the experimental data. The best distinction can be made in the 60-nm QW trace, where there are deep commensurability minima. The positions of these minima nearly match the spin-polarized marks with $n^*=n/3$. The agreement with the minority-density model is slightly better compared to the fixed-density model, as the former model captures the observed asymmetry of the minima around $\nu=3/2$. If CFs are assumed to be formed by holes in the $|0\downarrow\rangle$ LL (hole-density model), the predicted asymmetry in the expected positions of the commensurability minima would be in the wrong direction, i.e. the features on the $B^*>0$ are further away than those for $B^*<0$ (see Table I), which is inconsistent with the data. 

In general, the traces in Fig. 2 show that the positions of the commensurablity minima depend on the QW width and the density. As $W$ decreases, they change from a value nearly consistent with fully spin-polarized CFs to a value suggesting nearly spin-unpolarized CFs. For example, as $W$ decreases from 60 nm to 50 nm, at $n \simeq 1.8 \times 10^{11}$ cm$^{-2}$, the observed minima move closer to $\nu=3/2$. In the 50-nm-QW trace, they are nearly half-way between the two (solid and dashed) sets of vertical lines, indicating partial spin-polarization of the CFs. When $W=40$ nm and $n \simeq 1.78 \times 10^{11}$ cm$^{-2}$, the features are even closer to $\nu=3/2$ while for the $W=30$ nm QW (lowest trace in Fig. 2), the CFs appear to be nearly spin-unpolarized as the commensurability features are observed in close proximity to the dashed vertical lines, and certainly far from the solid vertical lines. 

In addition to the change in the positions of the commensurability features as $W$ varies, we can observe the effect of increasing density. At a fixed QW width, $W=40$ nm, but at higher density, $n \simeq 2.71 \times 10^{11}$ cm$^{-2}$, the positions of the resistance features move further away from $\nu=3/2$ compared to the lower density $W=40$ nm sample, as expected from Eq. (1). More important, the observed features become pronounced resistance minima and their positions are consistent with the predicted positions based on $f=1$, implying the CFs become more spin-polarized at higher density, as expected. 

\section{V. Spin-polarization of $\nu=3/2$ composite fermions}

Next, we attempt to quantitatively link the positions of the commensurability features for different QWs and electron densities to the CF spin-polarization. The trend is summarized in Fig. 4 where we show, for each commensurability feature, the ratio of its \textit{observed} position to the \textit{expected} position for spin-polarized CFs $(f=1)$ based on each of the three CF density models (Table I). In the fixed-density model, Fig. 4(a), the data from the $W=60$ nm and the higher density $40$ nm samples suggest that the CFs are nearly fully spin-polarized. The $B^*>0$ side (solid triangles) shows a slightly higher degree of spin-polarization which is expected because of the higher total field there. In the minority-density model, Fig. 4(b), the CFs in the $W=60$ nm QW and the higher density 40 nm QW are essentially fully spin-polarized, and the difference in the degree of spin-polarization on the two sides of $\nu=3/2$ is very small. As we discuss later, spin-transition data show that these two samples should exhibit full spin-polarization. Hence, the minority-density model better predicts the positions of the commensurability minima compared to the fixed-density model. In contrast, if we assume the CFs are formed by holes (Fig. 4 (c)), the polarization of the $B^*<0$ side is larger compared to the $B^*>0$ side. This is unreasonable since the applied magnetic field at $B^*>0$ is larger. 

If we assume the commensurability features represent the majority-spin carriers, the data from the $W=50$ nm QW suggest a slightly smaller degree of CF spin-polarization compared to the 60-nm-QW (see Figs. 4(a) and (b)). Data from the narrower wells at density $n \simeq 1.8 \times 10^{11}$ cm$^{-2}$ show that the CFs are even less spin-polarized. The error bars increase with decreasing QW width because the commensurability minima become less well-defined features. 

Regardless of what density model we use, the CFs become more spin-polarized with increasing well width and electron density. Even though the precise treatment of $n^*$ does not change our conclusions on the $W$- and $n$-dependence of the CF spin-polarization, it reveals subtle differences between the three density models. For the fixed-density and the minority-density models, the degree of spin-polarization is always higher for the $B^*>0$ side, which is expected since the minima there occur at a higher applied field. When $n^*$ is taken to be the density of holes, this trend is reversed, and the CFs on the $B^*<0$ side appear nearly fully spin-polarized while the degree of spin-polarization on the $B^*>0$ side is much lower. This is physically unreasonable since at lower applied field the Zeeman energy is lower. The minority-density model results in a higher degree of spin-polarization on both sides of $\nu=3/2$ and further decreases the difference between the two sides. This is reasonable in light of the fact that the difference in the applied field is $\simeq 0.5$ T, only about 10\% of the total applied field.

\begin{figure*}[!t]
\includegraphics[trim=0.8cm 0.2cm 3.0cm 0.2cm, clip=true, width=0.85\textwidth]{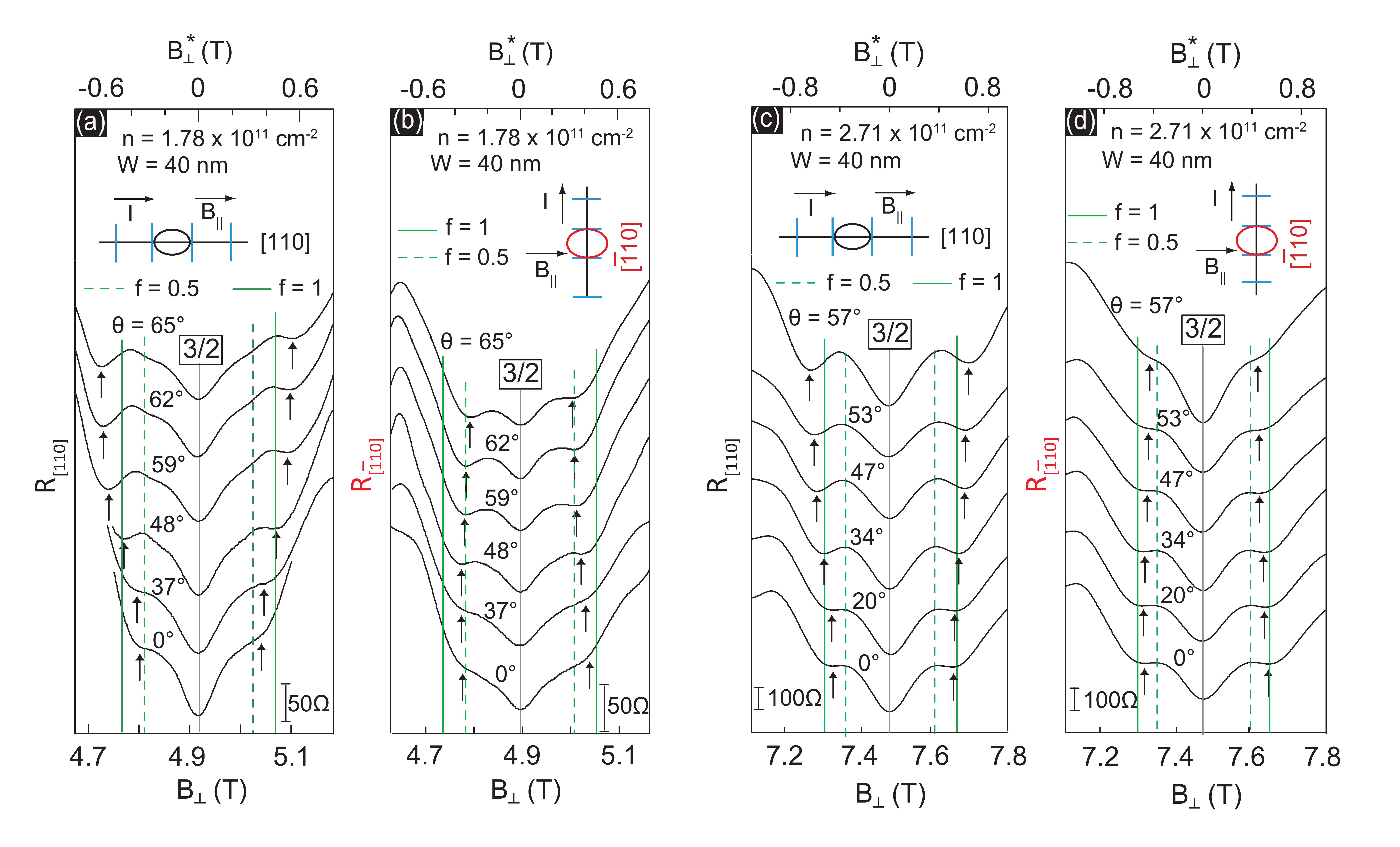}
\caption{\label{fig:Fig6} (color online) Magnetoresistance traces from the 40-nm-wide QWs: (a) and (b) $n \simeq 1.78$, (c) and (d) $2.71 \times 10^{11}$ cm$^{-2}$ taken along the $[110]$ and $[\overline{1}10]$ directions. For each set of data, solid $(f=1)$ and dashed $(f=0.5)$ vertical lines near $\nu=3/2$ mark the expected positions of the CF commensurability minima according to Eq. (1) with CF density $n^*$ equal to the minority density (see Table I). Vertical arrows mark the observed position of the commensurability features. }
\end{figure*}

We can recast our data in terms of the ratio of the Zeeman to the Coulomb energies, $E_Z/E_C$, which is independent of how we treat the CF density, and compare them to spin-polarization data from the CF spin transition measurements \cite{Du.PRL.1995,Yang.PRB.2014}, as shown in Fig. 5. The CF spin transitions, observed as a weakening and strengthening of the fractional quantum Hall states around $\nu=3/2$, provide sufficient information to depict the boundary beyond which the CFs are fully spin-polarized \cite{Du.PRL.1995,Yang.PRB.2014}. In Fig. 5, this expected boundary is drawn with solid lines for each QW. The positions of the lines are interpolated from the spin transition data, reported for $W=31, 42$ and 65 nm QWs \cite{Yang.PRB.2014}. The symbols in each graph of Fig. 5 are the calculated ratios $E_Z/E_C$ for the positions of the commensurability features/minima we observe on each side of $\nu=3/2$. The CFs are fully spin-polarized only in the 60 nm and the high density $W=40$ nm QWs. For all other samples, they are below the boundary for full spin-polarization.

The plots in Fig. 5 agree very well with the spin-polarization deduced from the plot of Fig. 4(b) which assumes a minority-density model for the CFs. In both Fig. 4(b) and Fig. 5, the CFs in the 60 nm and the high density $W=40$ nm QWs show nearly full spin-polarization; note that $B^*_{\text{min}}/B^*_{\text{expected, f=1}}$ values in Fig. 4(b) for these samples exceed $\simeq 0.95$. The fixed-density and the hole-density models, Figs. 4(a) and (c), on the other hand, predict a somewhat lower degree of spin-polarization. Hence, the comparisons in Fig. 5, which provide an independent check on the CF spin-polarization, further justify the use of the minority-density model for $n^*$. 

The strength of the commensurability features seen in Fig. 2 depends on both $W$ and $n$. The higher density 40 nm QW, as well as the 50 nm and 60 nm QWs, show pronounced resistance minima. In contrast, the CF commensurability in the 30 nm and 40 nm QWs is signalled only by weak features. This dependence and the sensitivity of our measurements to the majority-spin CF Fermi contour \cite{majority_spin_CF} could also be explained with the degree of CF spin-polarization. When the CFs are fully or nearly fully spin-polarized, the commensurability minima are well pronounced because there is essentially a single Fermi sea with a Fermi contour corresponding to CF density $n^*\simeq n/3$. In contrast, in the narrower QWs, where the CFs are partially spin-polarized, there are two Fermi contours with different sizes which should result in two nearby sets of commensurability features. Evidently, our measurements cannot resolve such features.

\section{VI. Effect of an in-plane magnetic field $B_{||}$}

The effect of an in-plane magnetic field $B_{||}$ on the CF commensurability minima is five--fold. First, $B_{||}$ couples to the CFs' out-of-plane orbital motion and makes their Fermi contours anisotropic. Second, the coupling also changes the charge distribution which in turn affects $E_C$. Third, $B_{||}$ increases the degree of CF spin-polarization because it raises $E_Z$ \cite{Ez}. Forth, when $B_{||}$ is sufficiently strong, there is a LL crossing between the $|0\downarrow\rangle$ and $|1\uparrow\rangle$ LLs as the separation between the symmetric and anti-symmetric LLs becomes smaller than $E_Z$. Finally, at even higher values of $B_{||}$, the system becomes bilayer-like. We first discuss the change in the positions of the CF commensurability minima as we tilt the sample in magnetic field at an angle $\theta$, then give an example of a LL crossing at high values of $\theta$, and quantify the CF Fermi contour anisotropy which we later use to draw conclusions about the spin-polarization as a function of the total magnetic field (see Fig. 1 inset for the experimental setup).

In the presence of $B_{||}$, the CF commensurability minima shift from their $\theta=0^{\circ}$ positions. The results for the 40-nm-wide QWs are summarized in Fig. 6. In the lower density 40 nm QW (Figs. 6(a) and (b)) at $\theta=0^{\circ}$, the CF commensurability is signaled by weak features rather than minima. The features fall between the expected values for spin-unpolarized (dashed green lines) and those for spin-polarized CFs (solid green lines). The CFs in the high-density 40-nm-wide QW are essentially fully spin-polarized even at $\theta=0^{\circ}$ (bottom traces in Figs. 6(c) and (d)). As the samples are tilted, in the $[110]$ direction (Figs. 6(a) and (c)), the positions of the observed resistance features move away from $\nu=3/2$. At the highest $\theta$, in both samples they are well outside the vertical solid green lines in Figs. 6(a) and (c). In the $[\overline{1}10]$ direction, shown in Figs. 6(b) and (d), the relative shift of the resistance minima is toward $\nu=3/2$. In addition, with increasing $B_{||}$, the features in the $n=1.78 \times 10^{11}$ cm$^{-2}$ sample gradually develop into well-defined resistance minima (Figs. 6(a) and (b)). We attribute this strengthening to the increase in the CF spin-polarization with tilt, similar to the case of the $W$- and $n$-dependence of the strength of the features.  The described shifts of the positions of $B^*_{\perp}$ with $B_{||}$ represent a general trend. In the $W \geq 50$ nm samples, the commensurability features remain well defined minima for a range of tilts and exhibit similar shifts with increasing $B_{||}$ (see, e.g., Fig. 7). Eventually, at high $\theta$, the system undergoes a LL crossing which destroys the commensurability features. 

\begin{figure}[!t]
\includegraphics[trim=0.5cm 2.5cm 0.cm 0cm, clip=true, width=0.420\textwidth]{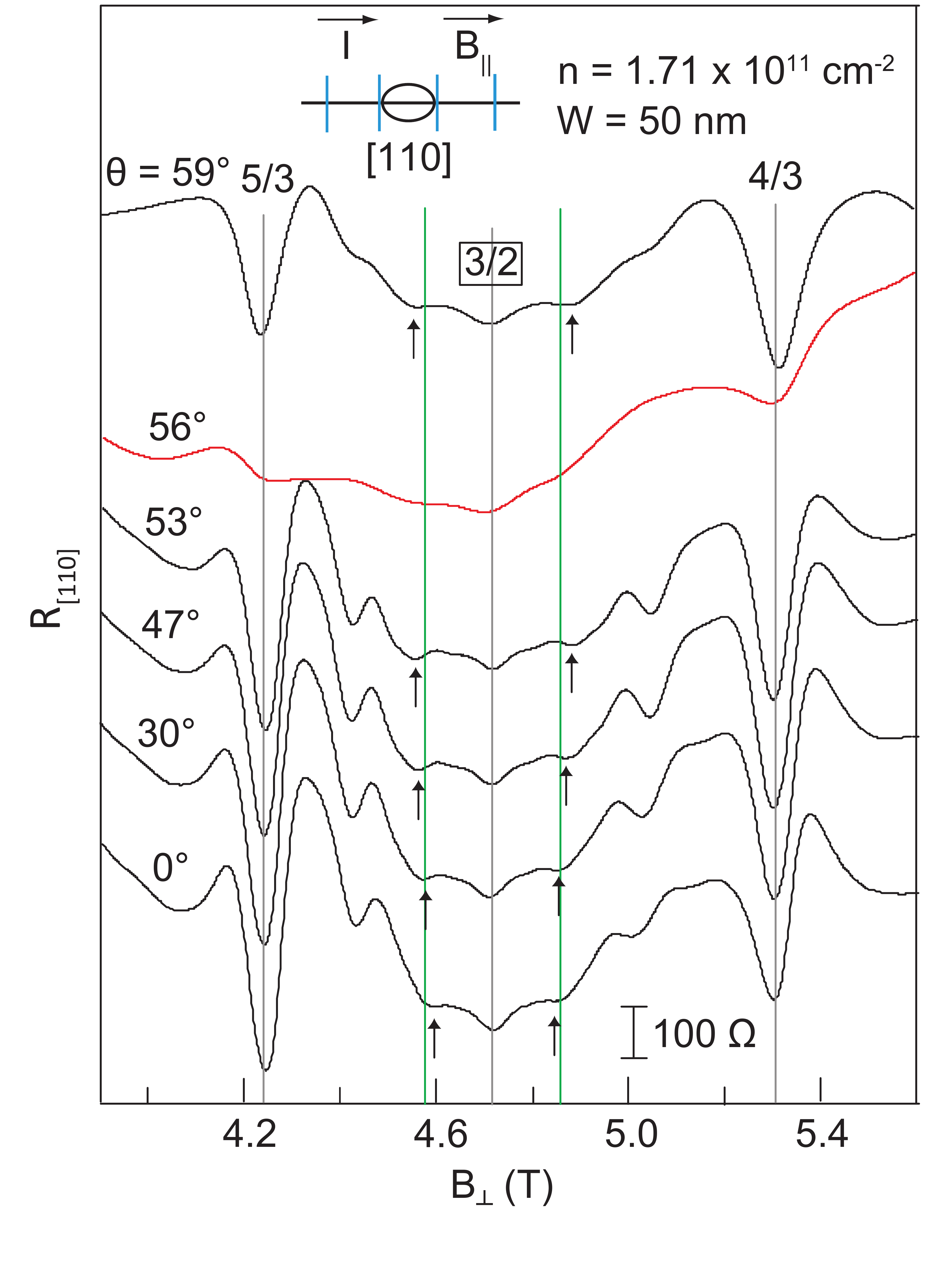}
\caption{\label{fig:Fig7} (color online) Magnetoresistance traces at several tilt angles $\theta$ from the 50-nm-wide QWs with $n \simeq 1.71 \times 10^{11}$ cm$^{-2}$ taken along the $[110]$ direction. The red trace ($\theta=56^{\circ}$) shows a weakening of all features observed near $\nu=3/2$, including the FQHSs at $\nu=4/3$ and 5/3 as well as the CF commensurability minima. It marks a crossing of the $|0\downarrow\rangle$ and $|1\uparrow\rangle$ LLs. Vertical arrows mark the observed positions of the commensurability features. The green vertical lines mark the expected positions of the CF commensurability minima according to Eq. (1) with CF density $n^*$ equal to the minority density (see Table I) and $f=1$ (signifying full spin-polarization).}
\end{figure}

In Fig. 7 we show an example of a LL crossing which appears at $\theta \simeq 56^{\circ}$ in the 50-nm-QW. The bottom trace in Fig. 7 is the same one shown in Fig. 2, where the positions of the commensurability minima indicate nearly full CF spin-polarization, consistent with the plots in Figs. 4 and 5. The red trace ($\theta=56^{\circ}$) shows the weakening of all features observed near $\nu=3/2$, including the FQHSs at $\nu=4/3$ and 5/3 as well as the CF commensurability minima. It marks a LL crossing between the $|0\downarrow\rangle$ (symmetric) and $|1\uparrow\rangle$ (antisymmetric) LLs. After the crossing, at even higher tilts, the Fermi level at $\nu=3/2$ is in the $|1\uparrow\rangle$ LL so that CFs are formed in an antisymmetric LL. Surprisingly, the CF commensurability features are restored. At even higher $\theta$ (data not shown) the features disappear again as the system becomes bilayer-like.

\begin{figure*}[t!]
\includegraphics[trim=0cm 1.0cm 0.0cm 0cm, clip=true, width=0.95\textwidth]{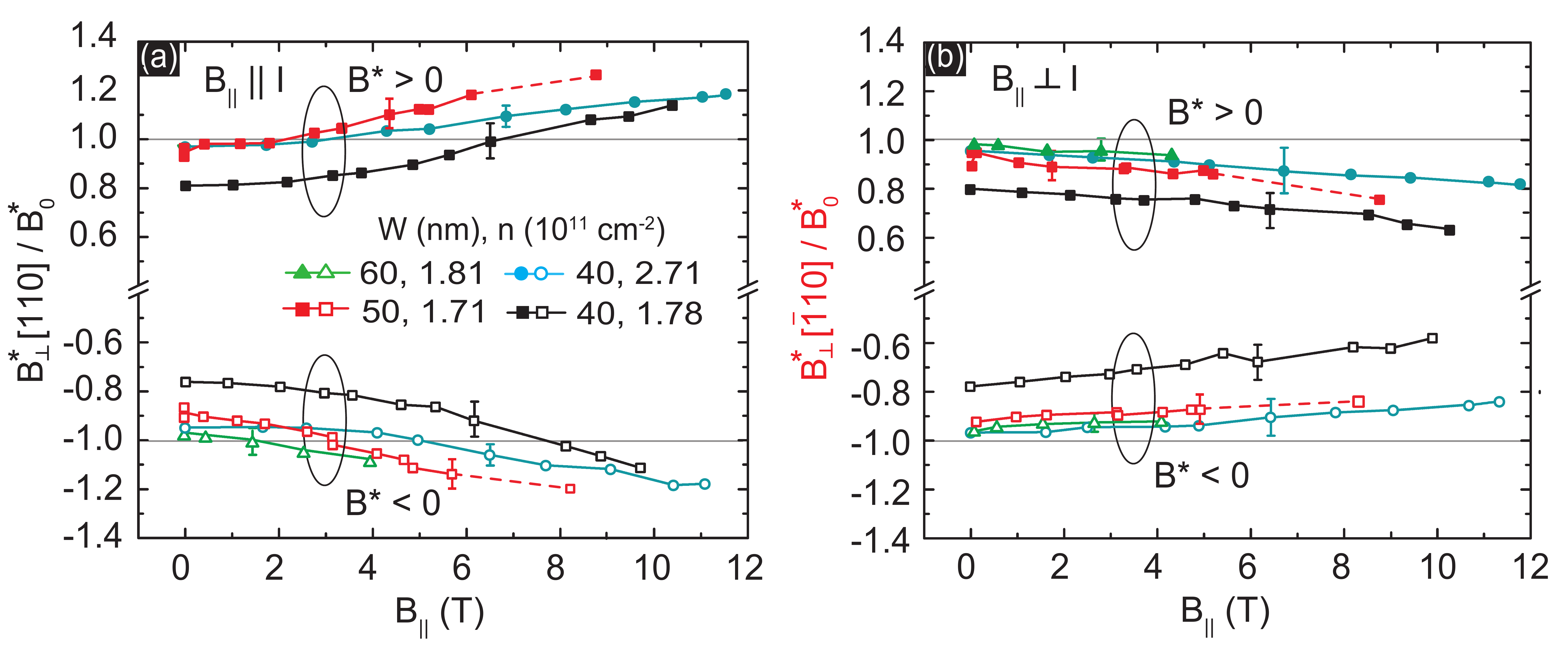}
\caption{\label{fig:Fig8} (color online) Positions of the resistivity minima for $W=40,50,60$ nm QWs measured along (a) $[110]$ and (b) $[\overline{1}10]$, normalized to $B^*_0$, the expected value from Eq. (1) a circular Fermi contour with $i=1, f=1$, and $n^*$ equal to the minority density. The solid lines are guides to the eye. Dashed lines for the $W=50$ nm QW represent the LL crossing region from Fig. 7, where the CF commensurability minima disappear. Representative error bars for each set of data points are also shown. The horizontal grey lines correspond to full spin-polarization $(f=1)$ and circular CF Fermi contour. }
\end{figure*}

Having described the qualitative change in the observed positions of the CF commensurability features/minima and the additional complications at high $B_{||}$, such as LL crossings and bilayerness, in Fig. 8 we study the effect of $B_{||}$ quantitatively in both arms of the L-shaped Hall bars. We summarize the change in the observed positions of the features by normalizing them to the expected values ($B^*_0$) for fully spin-polarized CFs $(f=1)$ with $n^*$ equal to the minority-density carriers in the $|0\downarrow\rangle$ LL and circular Fermi contour for each sample (Table I). In Figs.\;8(a) and (b), the normalized values in the two directions are labeled $B^*_{\perp}[110]$ and $B^*_{\perp}[\overline{1}10]$, respectively. The starting points of the $W=60$ nm and the higher density 40-nm-QW show nearly full spin-polarization, consistent with Figs. 4 and 5 \cite{footnote60nm}. The other samples show partial spin-polarization at $B_{||}=0$. For all samples, as $B_{||}$ increases, the values of $B^*_{\perp}[110]$ increase, while those for $B^*_{\perp}[\overline{1}10]$ decrease. We note that extracting the CF spin-polarization cannot be done directly from the data when $B_{||}$ is finite because the shift is also affected by the $B_{||}$-induced CF Fermi contour anisotropy. In the remainder of the paper we first discuss the Fermi contour anisotropy and then return to the question of spin-polarization. 

The evolution of the data points with increasing $B_{||}$ in Fig. 8 provides direct information about the anisotropy of the CF Fermi contours. In Fig. 9(a), we summarize the anisotropy data obtained by dividing the (interpolated) values of $B^*_{\perp}[110]$ by $B^*_{\perp}[\overline{1}10]$ and plotting the ratio as a function of $B_{||}$. All data points are from the $B^*>0$ side except for the $W=60$ nm points which are taken from the $B^*<0$ side of $\nu=3/2$ \cite{footnote60nm}. The dashed line for the $W=50$ nm QW is interpolated across the LL crossing. The CF Fermi contour distortion is significant for all samples. At low values of $B_{||}$, however, the proximity of the different curves and the large error bars make it difficult to extract the exact well-width dependence of the anisotropy. At higher $B_{||}$ a density dependence emerges in the 40 nm QWs: at higher density, $n = 2.71 \times 10^{11}$ cm$^{-2}$, the CF Fermi contours are less distorted compared to the $n=1.78 \times 10^{11}$ cm$^{-2}$ case. This can be understood in terms of a heavier CF effective mass at higher density, as we discuss below.  The $\nu=3/2$ Fermi contour anisotropy is much stronger than the distortion observed for CFs near $\nu=1/2$ at comparable values of $B_{||}$ \cite{eCFanisotropy.Kamburov.2012}. As we elaborate in the next paragraphs, we attribute this to the smaller CF effective mass near $\nu=3/2$ compared to the $\nu=1/2$ CFs.  

The dependence of the $\nu=3/2$ CF Fermi contour anisotropy on the CF effective mass and on the finite thickness of the conducting layer can be understood more quantitatively in a simple model, in the spirit of Fermi liquid theory \cite{Kamburov2.PRB.2014}. In this model, which provides an estimate for the expected CF Fermi contour anisotropy, an in-plane magnetic field $B_{||}$ couples the unbound in-plane motion to the quantized perpendicular motion of the quasi-2D carriers. We can evaluate its effect by treating $B_{||}$ in lowest-order perturbation theory \cite{ste68}. The model also assumes a parabolic dispersion with effective mass $m_\|$ for the in-plane motion and subband levels $E_n = \hbar^2 \pi^2 n^2 / (2m_z \, W^2)$ for the quantized perpendicular motion in a QW of width $W$ and out-of-plane mass $m_z$. Assuming fully spin-polarized electrons, this yields an elliptic Fermi contour with minor and major radii \cite{error}: 
\begin{equation}
\label{eq:model}
k_{\mp} = \sqrt{ 4 \pi n} \left( 1 - \frac{2^{10}}{3^5 \pi^6} \frac{e^2 \, B_{||}^2}{\hbar^2} \frac{W^4 \, m_z}{m_\|} \right)^{\pm 1/4}.
\end{equation} 
This perturbative expansion in $B_{||}$ is valid as long as the distortion of the Fermi contour described by the second term in the round brackets remains small. Here we may expect that the physics of CFs characterizes the in-plane dynamics of the quasi-particles in our experiments, implying $m_\|$ should be approximately the effective mass of CFs that contains electron-electron interactions. The quantized perpendicular motion of the quasi-particles giving rise to the formation of electric subbands should reflect, on the other hand, the band dynamics which is characterized by the band mass of electrons in GaAs \cite{Winkler.Book.2003}.

\begin{figure*}[!t]
\includegraphics[trim=0.cm 0.5cm 0.cm 0.5cm, clip=true, width=0.85\textwidth]{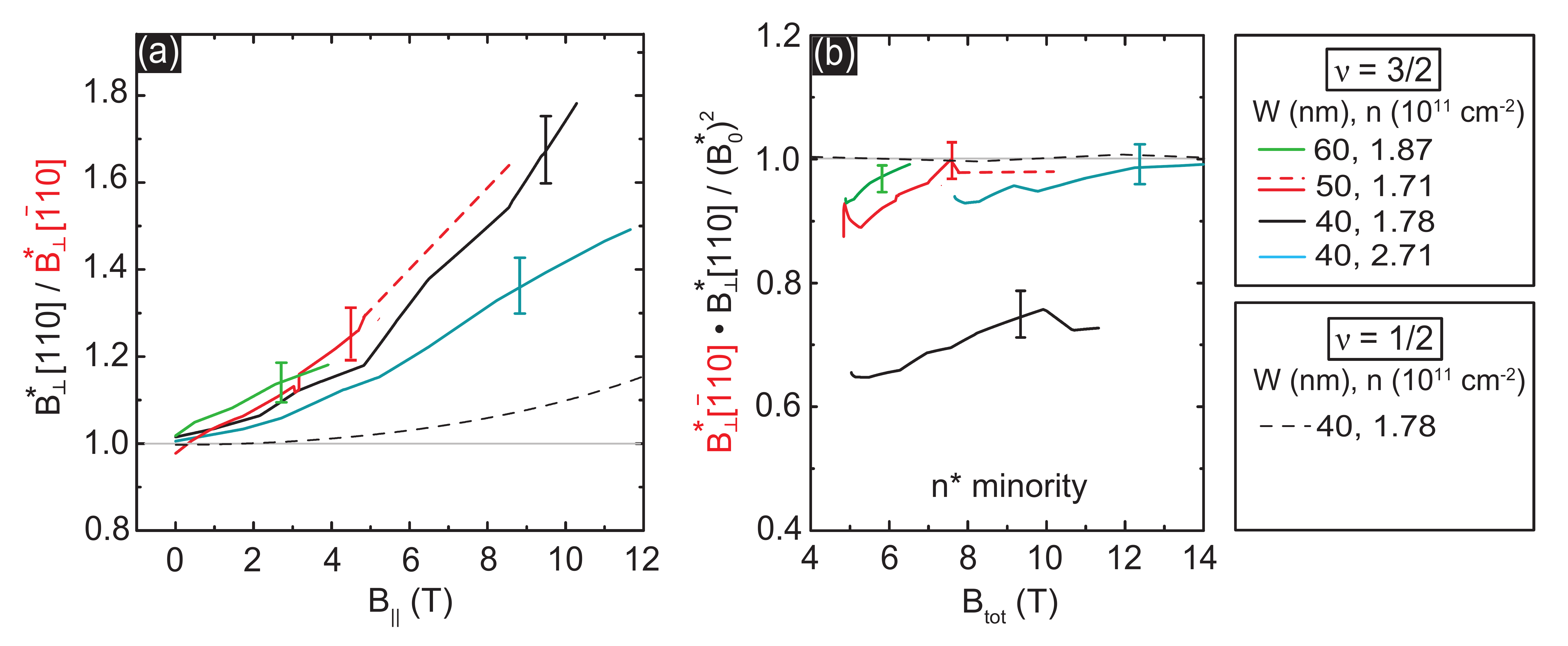}
\caption{\label{fig:Fig9} (color online) (a) Solid lines: Relative change in the positions of the resistivity minima along the two perpendicular directions obtained by dividing the interpolated values of $B^*_{\perp}[110]$ and $B^*_{\perp}[\overline{1}10]$ from Fig. 8. The values shown are from the $B^*>0$ side except for the $W=60$ nm QW, where the data are from the $B^*<0$ side. Dashed portion of the red curve is based on interpolation across the LL crossing region in the 50 nm QW. Dashed black curve: Anisotropy of the CF Fermi contour near $\nu=1/2$ measured in the lower density $W=40$ nm QW \cite{eCFanisotropy.Kamburov.2012}. (b) Composite fermion spin-polarization expressed by treating the normalized observed positions of the commensurability minima using $n^*$ corresponding to the minority carrier density. Data for $\nu=1/2$ CFs are shown with a dashed black curve. }
\end{figure*}

The CF effective mass $m^{CF}_{||}$ is in general very different from the electron band effective mass in GaAs, $m_e=0.067$ (in units of the free-electron mass), and depends on interaction which increases as $\sqrt{B_{\perp}}$ \cite{Jain.2007}. Near $\nu=1/2$, for example, $m^{CF}_{||}$ is measured to be $ \simeq 0.6$ at $B_{\perp} \simeq 9$ T \cite{Du.PRL.1993}, while in the same sample the measured $m^{CF}_{||}$ for CFs near $\nu=3/2$ at $B_{\perp} \simeq 3$ T is $\simeq 0.4$ \cite{Du.SS.1996}. 

In light of its dependence on $m_{||}$ and QW width, the theoretical model based on Eq. (2) has three qualitative implications for the Fermi contour anisotropy of the various particles in our measurements. First, it predicts that, for a given sample and density, the electrons near $B_{\perp} = 0$ would exhibit the highest Fermi contour anisotropy because they have the smallest $m_{||}$ while the $\nu=1/2$ CFs would show the least anisotropy. The anisotropy for the $\nu=3/2$ CFs would be smaller than the anisotropy for electrons but larger than that of the $\nu=1/2$ CFs. This is indeed the qualitative trend seen in our results. For example, for the 40-nm-wide QW sample with density $1.78 \times 10^{11}$ cm$^{-2}$ and at $B_{||} \simeq 10$ T, the anisotropy is $\simeq 2.3$ for electrons \cite{Kamburov2.PRB.2014}, $\simeq 1.7$ for $\nu=3/2$ CFs (black curve in Fig. 9(a)), and only about 1.1 for $\nu=1/2$ CFs \cite{eCFanisotropy.Kamburov.2012} (also see the dashed gray curve in Fig. 9(a)). Second, the model of Eq. (1) predicts a QW-width dependence, which is also qualitatively observed: wider QWs show larger CF Fermi contour anisotropy. Third, the model explains the density dependence of the data for the two $W=40$ QW samples. In particular, $m^{CF}_{||} \propto \sqrt{B_{\perp}}$ implies a larger $m^{CF}_{||}$ at higher density, resulting in a smaller CF Fermi anisotropy as seen in Fig. 9(a). We emphasize that we only see a qualitative agreement between the trends predicted by Eq. (2) and those observed experimentally; this is not surprising since the model is only perturbative and not rigorous. 

As mentioned above, another consequence of the application of $B_{||}$ is the enhancement of $E_Z$ with respect to $E_C$ and the increase in the degree of CF spin-polarization. To quantitatively assess the change in the CF spin-polarization, we plot in Fig. 9(b) the normalized product of the observed field positions of the CF commensurability resistance minima ($B^*_{\perp}[110]$ and $B^*_{\perp}[\overline{1}10]$) as a function of total magnetic field $B_{\text{tot}}$. The CF density is taken to be the minority-density. The dashed section of the 50-nm QW trace represents the LL crossing region in Figs. 7 and 8. We also show, with a dashed black curve, the $\nu=1/2$ data from the $W=40$ nm QW with $n=1.78 \times 10^{11}$ cm$^{-2}$ \cite{eCFanisotropy.Kamburov.2012}.

The CFs in our $\nu=1/2$ study are fully spin-polarized so the $\nu=1/2$ data appear as a nearly flat line (see  dashed black curve in Fig. 9(b)), which in turn allows us to conclude that the CF Fermi contours remain elliptical even at high values of $B_{||}$. Assuming the $\nu=3/2$ CF Fermi contours also remain nearly elliptical, Fig. 9(b) is effectively a plot of the CF spin-polarization \cite{polarization}. Indeed, all samples except for the lower-density $W=40$ nm QW appear to reach full spin-polarization. However, the apparent spin-polarization in the lower-density $W=40$ nm sample remains surprisingly small. Even at $B_{||}=11$ T, the CFs there are still partially spin-polarized which is not consistent with the other samples. It is likely that because of the significant anisotropy at high $B_{||}$ (Fig. 9(a)), the CF Fermi contour is no longer elliptical which prevents use from accurately determining the CF spin-polarization from a plot like the one in Fig. 9(b).  

\section{VII. Summary and Conclusions}

In summary, we observe pronounced commensurability minima in the vicinity of $\nu=3/2$. Their positions with respect to the field at $\nu=3/2$ are slightly asymmetric. The presence of this asymmetry necessitates the proper treatment of the CF density $n^*$ and reveals that the CFs near $\nu=3/2$ are likely formed by the minority carriers in the $|0\downarrow\rangle$ LL, in agreement with the observations for $\nu=1/2$ \cite{Kamburov.2014}. By deducing the size of the CF Fermi wave vector, we find that the CF spin-polarization depends on the QW width and the electron density. We attribute this dependence to the change in the ratio $E_Z/E_C$ of the Zeeman and Coulomb energies: $E_C$ decreases with increasing layer thickness as the Coulomb interaction softens, while $E_Z$ is enhanced at higher densities with respect to $E_C$. In the presence of $B_{||}$, the CF Fermi contours become distorted and the CF spin-polarization increases. Qualitatively, the $B_{||}$-induced CF Fermi contour anisotropy seen near $\nu=3/2$ is larger than the distortions observed near $\nu=1/2$ because of the smaller CF in-plane mass near $\nu=3/2$. Quantitatively, however, the $\nu=3/2$ CF Fermi contour distortion is stronger than our expectations based on a simple perturbative model. We hope our experimental results would stimulate a more rigorous theoretical treatment of the CF Fermi contour anisotropy.

\begin{acknowledgments}
We acknowledge support through the DOE BES (DE-FG02-00-ER45841) for measurements, and the Gordon and Betty Moore Foundation (Grant GBMF4420), Keck Foundation, NSF (ECCS-1001719, DMR-1305691, and MRSEC DMR-0819860) for sample fabrication and characterization. A portion of this work was performed at the National High Magnetic Field Laboratory which is supported by National Science Foundation Cooperative Agreement No. DMR-1157490, the State of Florida and the US Department of Energy. Work at Argonne was supported by DOE BES under Contract No. DE-AC02-06CH11357. We thank S. Hannahs, T. Murphy, and A. Suslov at NHMFL for valuable technical support during the measurements. We also express gratitude to Tokoyama Corporation for supplying the negative e-beam resist \mbox{TEBN-1} used to make the samples.
\end{acknowledgments}


\begin{thebibliography}{99}
\bibitem{gfactor} Throughout this paper we use $\epsilon=13.1\epsilon_0$ and $|g|=0.44$ for GaAs.
\bibitem{majority_spin_CF} Partial spin-polarization of the CFs implies that two Fermi contours should be visible in commensurability experiments. Our measurements appear to be sensitive only to the majority-spin Fermi contour.
\bibitem{Ez} The in- and out-of-plane components of the electron $g$-factor can be different. A detailed assessment of $g$-factor anisotropy and comparison with theory is beyond the scope of this work.
\bibitem{footnote60nm} The quality of the $B^*>0$ data from the 60 nm sample along [110] is not sufficiently good to accurately yield the positions of the CF commensurability minima.
\bibitem{polarization} Figure 9(b) approximates the CF spin-polarization well but is not exact because, at a given value of total field, the commensurability minima corresponding to the major axes of an elliptical Fermi contour would occur at different values of $B^*_{\perp}$. In the minority-density model $n^*$ is not a fixed quantity and varies with $B^*_{\perp}$.
\end{thebibliography}
\end{document}